\documentclass[twocolumn,showpacs,aps]{revtex4}%
\usepackage{amsfonts}
\usepackage{amsmath}
\usepackage{amssymb}
\usepackage{graphicx}
\providecommand{\U}[1]{\protect\rule{.1in}{.1in}}

\begin{document}
\title{Nonlinear Rashba Model and Spin Relaxation in Quantum Wells}
\author{W. Yang}
\author{Kai Chang}
\altaffiliation[Author to whom the correspondence should be addressed. Electronic address: ]{kchang@red.semi.ac.cn}

\affiliation{SKLSM, Institute of Semiconductors, Chinese Academy of Sciences, P. O. Box
912, Beijing 100083, China}


\begin{abstract}
We find that the Rashba spin splitting is intrinsically a nonlinear function
of the momentum, and the linear Rasha model may overestimate it significantly,
especially in narrow-gap materials. A nonlinear Rashba model is proposed,
which is in good agreement with the numerical results from the eight-band
\textbf{k}$\cdot$\textbf{p} theory. Using this model, we find pronounced
suppression of the D'yakonov-Perel' spin relaxation rate at large electron
densities, and a non-monotonic dependence of the resonance peak position of
electron spin lifetime on the electron density in [111]-oriented quantum
wells, both in qualitative disagreement with the predictions of the linear
Rashba model.

\end{abstract}
\maketitle

Recently there has been growing interest in the field of \textit{Spintronics}%
\cite{Prinz,Wolf,DasSarma}, which explores the electron spin, in addition to
the electron charge, to realize new functionalities in future electronic
devices.\cite{Datta,Nitta1,Loss} A promising approach implementing such
spintronic devices is to utilize the Rashba spin-orbit interaction caused by
structure inversion asymmetry in quantum wells (QW's),\cite{Rashba} which can
be controlled by gate voltages\cite{Nitta2,Grundler} as well as band structure
engineering.\cite{book,Pfeffer2} Approximate analytical expressions based on
second- or third-order perturbation theories\cite{book,Pfeffer2,Bassani}
suggest that the Rashba spin splitting (RSS) is a linear function of the
in-plane wave vector $k_{\parallel}$. This linear Rashba model has been widely
used to investigate the various spin-related properties of low-dimensional
semiconductor structures, e.g., electron spin
relaxation\cite{Averkiev,SR1,SR2,YCChang,Kainz} and the newly discovered spin
Hall effect\cite{SHE1,SHE2,SHE3,SHE4,SHE5,SHE6,SHE7}. However, recent
numerical calculations\cite{Winkler,Lamari,Bassani} show that the RSS in
certain semiconductor QW's deviates from the linear behavior at large
$k_{\parallel}$, although the underlying physics remains unclear. Since the
linear Rashba model is still widely used by the mainstream researchers, it is
important to explore the underlying physics beneath such deviation and, if
necessary, check the validity of the linear Rashba model.

In this work, we find from analytical derivation from the eight-band
$\mathbf{k}\cdot\mathbf{p}$ theory that the RSS is intrinsically a nonlinear
function of the wave vector, which is caused by the weakening of the interband
coupling with increasing kinetic energy of the electron in the conduction
band. We show from numerical comparisons that the deviation of the linear
Rashba model could be surprisingly large at large wave vectors, especially for
narrow-gap materials. These facts substantiate the necessity for a nonlinear
Rashba model. We propose such a model, which is in good agreement with the
numerical results from the eight-band $\mathbf{k}\cdot\mathbf{p}$ theory for
various QW's. This nonlinear Rashba model would lead to a series of
significant modifications to the various spin-related properties of the
electron, most of which has been investigated based on the linear Rashba
model. For example, we find pronounced suppression of the D'yakonov-Perel'
(DP) spin relaxation rate (SRR) at large electron density, in qualitative
disagreement with the prediction of the linear Rashba
model.\cite{Averkiev,Kainz} The resonant enhancement of the electron spin
lifetime in [111]-oriented quantum wells\cite{YCChang} also exhibits
qualitatively different behavior from the linear Rashba model. The values of
the SRR obtained by the two models differ by up to several orders of
magnitude. Further surprising consequences are expected when this nonlinear
Rashba model is applied to other fields.

For [001]-oriented heterostructures, the eight-band Hamiltonian\cite{Burt}
$H=H_{k}+V$, where $V=eFz$ is the external electric field induced potential
and $H_{k}$ is the eight-band envelope function Hamiltonian. Neglecting the
off-diagonal elements in the valence bands and eliminating the valence band
components of the envelope function, the effective conduction band Hamiltonian
is obtained as%
\begin{equation}
H_{eff}(k_{\Vert})=E_{c}(z)+V(z)+\mathbf{k}\frac{\hbar^{2}}{2m^{\ast}%
}\mathbf{k}+\alpha_{0}(z)(k_{\Vert}\times e_{z})\cdot\sigma, \label{Hamil}%
\end{equation}
where $\alpha_{0}=\hbar^{2}/(6m_{0})\partial\gamma(z)/\partial z$,
$\gamma(z)=E_{P}[1/U_{lh}(z)-1/U_{SO}(z)]$. $E_{P}=2m_{0}P_{0}^{2}/\hbar^{2},$
and $m^{\ast}$ is the effective mass given by $m^{\ast}=m_{0}(\gamma
_{c}+2E_{P}/(3U_{lh})+E_{P}/(3U_{SO}))^{-1}$, $U_{lh}=E-H_{lh}$,
$U_{SO}=E-H_{SO}$, and $H_{lh}$ and $H_{SO}$ are the diagonal elements of the
light-hole and spin-orbit split-off bands in the eight-band Hamiltonian. From
Eq. (\ref{Hamil}), we find that the dominant contribution to the RSS consists
of the interface term $\Delta E_{n}^{(1)}$ in the valence band and the
external electric field term $\Delta E_{n}^{(2)}$,%
\begin{align}
\Delta E_{n}^{(1)}(k_{\parallel})  &  =\frac{\hbar^{2}}{3m_{0}}k_{\parallel
}\sum\limits_{j}\left\vert F_{n}(z_{j})\right\vert ^{2}[\gamma(z_{j}%
^{+})-\gamma(z_{j}^{-})],\label{RSSa}\\
\Delta E_{n}^{(2)}(k_{\parallel})  &  =\frac{\hbar^{2}}{3m_{0}}E_{P}%
eFk_{\parallel}\int dz\ \left\vert F_{n}(z)\right\vert ^{2}(U_{\text{lh}}%
^{-2}-U_{\text{so}}^{-2}). \label{RSSb}%
\end{align}
Here $F_{n}(z)$ is the envelope function of the $n$th subband along the $z$
axis, and $\{z_{j}\}$ denote the $z$ coordinates of the interfaces. According
to the Ehrenfest's theorem, the external electric field contribution $\Delta
E_{n}^{(2)}$ is approximately cancelled by the interface electric field
contribution in the conduction band.\cite{Winkler2} As a result, the dominant
contribution to RSS comes from the interface term only.\cite{Pfeffer2} From
Eq. (\ref{RSSa}), we see that RSS is approximately inversely proportional to
the effective band gap $E_{g}^{eff}=E-E_{v},$ where $E_{v}$ is the valence
band edge and $E$ is the electron energy. Since the electron energy $E$
increases approximately quadratically as $k_{\parallel}$ increases, the RSS
would always begin to decrease after $k_{\parallel}$ has exceeded a critical
value. This reveals that the previously found deviation of the RSS from linear
behavior in certain QW's\cite{Winkler,Bassani,Lamari} is actually a universal
behavior, suggesting that the widely used linear Rashba model need to be
checked and, if necessary, replaced by a nonlinear one.

Enlightened by above discussions, we propose the following two-coefficient
nonlinear Rashba model to describe the RSS of a given subband $n$,%

\begin{equation}
\Delta E_{n}(k_{\parallel})=\frac{2\alpha_{n}k_{\parallel}}{1+\beta
_{n}k_{\parallel}^{2}}, \label{model2}%
\end{equation}
where $\alpha_{n}$ is the linear Rashba coefficient of the $n$th subband for
the widely used linear Rashba model $\Delta E_{n}(k_{\parallel})=2\alpha
_{n}k_{\parallel}$, while $\beta_{n}k_{\parallel}^{2}$ describes the
contribution from the kinetic energy of electron in the $n$-th subband. The
latter leads to the decrease of RSS when $k_{\parallel}$ exceeds a critical
value $k_{0}=1/\sqrt{\beta_{n}}$, which is determined from $\left[  d(\Delta
E_{n}(k_{\parallel}))/dk_{\parallel}\right]  _{k_{\parallel}=k_{0}}=0$.

\begin{figure}[ptb]
\includegraphics[width=\columnwidth]{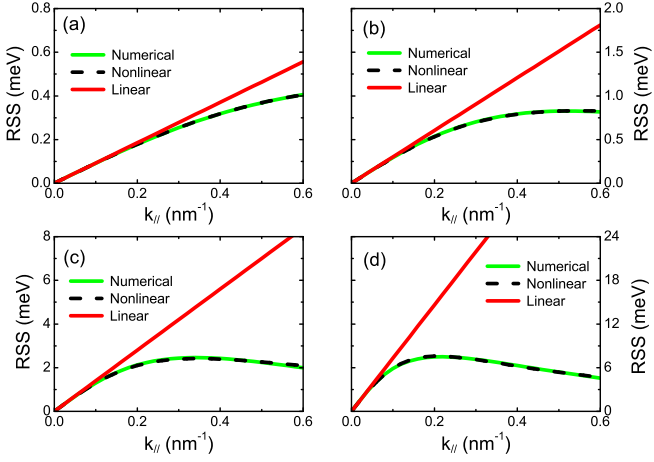}\caption{(Color online)
Comparison of the linear (red or dark gray lines) and nonlinear (black lines)
models with the numerical results (green or light gray lines) obtained from
the eight-band $\mathbf{k}\cdot\mathbf{p}$ theory for the lowest conduction
subband of 15-nm-wide (a) GaAs/Ga$_{0.67}$Al$_{0.33}$As, (b) Ga$_{0.47}%
$In$_{0.53}$As/Al$_{0.48}$In$_{0.52}$As, (c) InAs/In$_{0.75}$Al$_{0.25}$As,
and (d) Hg$_{0.79}$Cd$_{0.21}$Te/CdTe QW's with an electric field 60 kV/cm.}%
\end{figure}In Fig. 1, we compare the linear and nonlinear Rashba models with
the numerical results obtained from the eight-band $\mathbf{k}\cdot\mathbf{p}$
theory for different QW's. The band gap of the well material ranges from 1.519
eV (for GaAs) to 0.1 eV (for Hg$_{0.79}$Cd$_{0.21}$Te). We see that the
deviation of the linear model from the numerical results could be surprisingly
large, especially for narrow gap QW's, while the nonlinear model is in good
agreement with the numerical results. Thus we substantiate the necessity to
use the nonlinear Rashba model instead of the linear one.

To characterize the nonlinear Rashba model, we consider the dependence of the
Rashba coefficients $\alpha_{n}$ and $\beta_{n}$ on the various band
parameters. To the lowest order, the probability asymmetry at the two
interfaces of the QW is proportional to the external electric field $F$, then
Eq. (\ref{RSSa}) suggests $\Delta E_{n}(\mathbf{k}_{\parallel})\sim
k_{\parallel}F/[E_{g}+E_{n0}+\hbar^{2}k_{\parallel}^{2}/(2m_{n}^{\ast})], $
where $E_{n0}$ is the quantum confining energy and $m_{n}^{\ast}$ is the
effective mass of the $n$th subband. This shows that%

\begin{align}
\alpha_{n}  &  \sim\frac{F}{E_{g}+E_{n0}},\label{alpha}\\
\beta_{n}  &  \sim\frac{1}{m_{n}^{\ast}(E_{g}+E_{n0})}. \label{beta}%
\end{align}
$\alpha_{n}$ and $\beta_{n}$ both increase with decreasing band gap and
subband index $n$, or increasing well width, i.e, the quantum confining energy
$E_{n0}$. The significant difference is that $\alpha_{n}$ increases with
increasing external electric field $F$, while $\beta_{n}$ is approximately
independent of $F$. Notice, however, additional dependence of $\alpha_{n}$ and
$\beta_{n}$ on the external electric field may comes from the quantum
confining energy, especially at large electric fields.

\begin{figure}[ptb]
\includegraphics[width=\columnwidth]{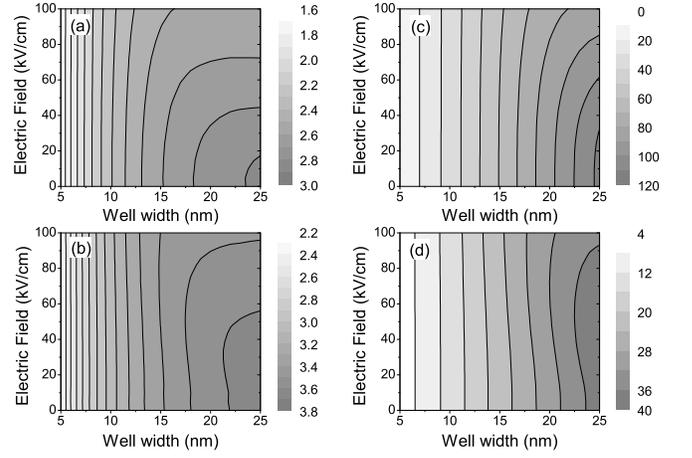}\caption{Rashba coefficient
$\alpha/F$ (upper panels) and $\beta$ (down panels) of the lowest conduction
subband as a function of the well width and external electric field for
Ga$_{0.47}$In$_{0.53}$As/Al$_{0.48}$In$_{0.52}$As [(a) and (b)], and
Hg$_{0.79}$Cd$_{0.21}$Te/CdTe [(c) and (d)] QW's. The units of $\alpha/F$ and
$\beta$ are meV nm/(100 kV/cm) and nm$^{2}$, respectively.}%
\end{figure}In Fig. 2, we plot the Rashba coefficient $\alpha/F$ and $\beta
$\ of the lowest conduction subband (obtained by fitting the results from the
eight-band $\mathbf{k}\cdot\mathbf{p}$ theory) as functions of the well width
and external electric field for different semiconductor QW's. Some common
features can be observed. Firstly, $\alpha/F$ and $\beta$\ are weakly
dependent on the external electric field. This shows that $\alpha$ is
approximately proportional to the external electric field, while $\beta$ is
approximately independent of $F$, in agreement with Eqs. (\ref{alpha}) and
(\ref{beta}). Secondly, $\alpha/F$ and $\beta$ increase with the increase of
the well width and begin to saturate at large well width. The saturation value
decreases with the increase of the electric field, and this trend becomes
increasingly pronounced when the band gap of the well decreases. The increase
and saturation behavior comes from the competition between the band gap
$E_{g}$ and the quantum confining energy $E_{n0}$ especially at small well
width. The decrease of the saturation values of $\alpha/F$ and $\beta$ at
large well width comes from the enhancement of the quantum confining energy
$E_{n0}$ due to the triangular potential induced by the external electric
field. It also reflects the fact that the relationship $\alpha\propto F$
begins to break down for wide QW's. Thirdly, the critical well width at which
$\alpha/F$ begins to saturate decreases with the increase of the electric
field. This can be understood from the competition between the square
potential [produced by the conduction band profile $E_{c}(z)$] and the
triangular potential (produced by the external electric field). With the
increase of the well width, the square potential becomes weaker and the
triangular potential begins to dominate. This transition occurs at a smaller
well width when the external electric field get stronger, leading to the
saturation of the quantum confining energy $E_{n0}$ and, consequently, the
Rashba coefficients $\alpha/F$ and $\beta$ at a smaller well width. The
similarities between $\alpha/F$ and $\beta$ are in agreement with our
analytical discussions in Eqs. (\ref{alpha}) and (\ref{beta}). However, we
should notice that $\alpha$ is approximately linearly dependent on the
external electric field $F$, while $\beta$ is approximately independent of
$F$.\begin{figure}[ptb]
\includegraphics[width=\columnwidth]{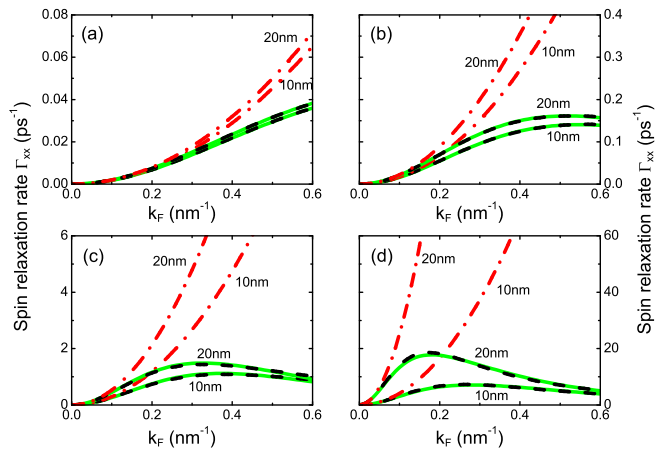}\caption{(Color online) Low
temperature DP SRR in the lowest conduction subband of 20nm- and 10nm-wide (a)
GaAs/Ga$_{0.67}$Al$_{0.33}$As, (b) Ga$_{0.47}$In$_{0.53}$As/Al$_{0.48}%
$In$_{0.52}$As, (c) InAs/In$_{0.75}$Al$_{0.25}$As, and (d) Hg$_{0.79}%
$Cd$_{0.21}$Te/CdTe QW's with an electric field 60 kV/cm and momentum
relaxation time 0.1 ps. The results are obtained from the eight-band model
(green or light gray lines), the nonlinear (black lines) and linear (red or
dark gray lines) Rashba models, respectively.}%
\end{figure}

In the above, we see that the nonlinear Rashba model differs significantly
from the linear model. As a result, we expect that this nonlinear Rashba model
would lead to a series of modifications to the various spin-related properties
of the electron, which have been investigated based on the linear Rashba
model.\cite{Averkiev,SR1,SR2,YCChang,Kainz,SHE1,SHE2,SHE3,SHE4,SHE5,SHE6,SHE7}
As an example, we consider the electron spin relaxation caused by Rashba
spin-orbit coupling in two-dimensional electron gas. The electron spin
relaxation process has received intensive interest recently because it plays
an essential role in the practical application of spintronic devices and
quantum information
processing.\cite{spinrelax1,spinrelax2,spinrelax3,Awschalom2} For electron in
n-doped semiconductors at large electron density, DP mechanism is the dominant
process and it has been investigated by several
groups\cite{Averkiev,YCChang,Kainz} using the linear Rashba model in the
framework of the density matrix formalism. We follow this procedure and
calculate the DP SRR [$\Gamma_{xx}=\Gamma_{yy}=\Gamma_{zz}/2$, $\Gamma
_{ij}=0\ (i\neq j)$] using the eight-band $\mathbf{k}\cdot\mathbf{p}$ theory,
the nonlinear and linear Rashba models, respectively. The results for
different QW's are shown in Fig. 3. The SRR's from the nonlinear Rashba model
show pronounced suppression at large Fermi wave vector $k_{F}$ (i.e., electron
density), in good agreement with the numerical results from the eight-band
$\mathbf{k}\cdot\mathbf{p}$ theory, while those from the linear Rashba model
increase monotonically with $k_{F}$, which qualitatively disagrees with the
nonlinear model. The difference between the linear and nonlinear Rashba models
increases significantly with the increase of $k_{F}$. Taking the 20 nm wide
QW's for example, at $k_{F}=0.4$ nm$^{-1}$ (corresponding electron density
$n_{e}\approx2.5\times10^{12}$ cm$^{-2}$, which is of interest to, e.g.,
transport measurements), the relative difference between the two models
reaches 40\%, 140\%, 500\%, and 3800\% for Figs. 3(a)-3(d), respectively.

\begin{figure}[ptb]
\includegraphics[width=\columnwidth]{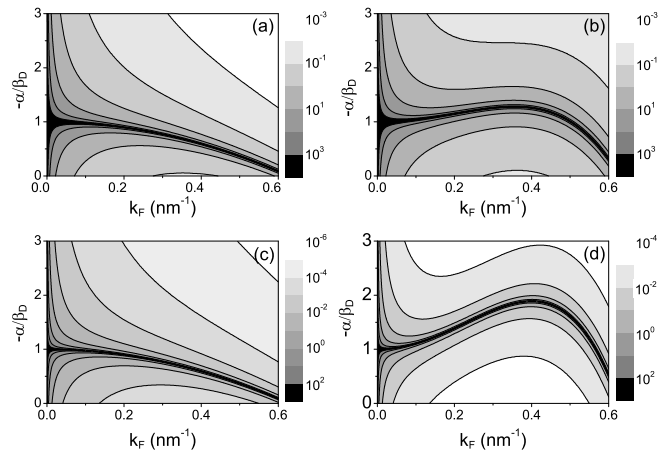}\caption{{}$\tau_{zz}$ (in units
of ps) for 10-nm-wide InAs/In$_{0.75}$Al$_{0.25}$As [(a) and (b)] and
Hg$_{0.79}$Cd$_{0.21}$Te/CdTe [(c) and (d)] QW's with a typical momentum
relaxation time 0.1 ps. The linear (nonlinear) Rashba model is used for
(a)\ and (c) [(b) and (d)].}%
\end{figure}In the above, we consider only the Rasbha spin-orbit coupling in
order to demonstrate the significant consequences that are introduced by our
nonlinear Rasbha model. To perform realistic calculations, however, both the
nonlinear Rashba model and the Dresselhaus spin-orbit coupling should be
included. Recently it was predicted based on the linear Rashba model that the
interplay between the Rashba and Dresselhaus spin-orbit interactions would
greatly enhance the electron spin lifetime in [111]-orientated quantum wells
when the linear Rashba coefficient $\alpha$ and the linear Dresselhaus
coefficient $\beta_{D}$ are opposite (i.e., when the ratio $-\alpha/\beta
_{D}=1$).\cite{YCChang} As a second example, we reconsider this problem using
the nonlinear Rashba model. The electron spin lifetime $\tau_{zz}$ (along the
growth direction of the QW) obtained from the two models are compared in Fig.
4 as functions of the ratio $(-\alpha/\beta_{D})$ and Fermi wave vector
$k_{F}$. The black regions correspond to the resonant enhancement regions of
the spin lifetime. The most striking feature is that the ratio $(-\alpha
/\beta_{D})$ at the resonance peak (latter referred to as the peak position)
shows non-motonic dependence on the Fermi wave vector [cf. Figs. 4(b) and
4(d)], which differs qualitatively from the monotonically decreasing behavior
of the linear Rashba model [cf. Figs. 4(a)\ and 4(c)]. Here the monotonic
decrease of the peak position with increasing $k_{F}$ in the linear Rashba
model comes from the Dresselhaus $k^{3}$ term, while the nonlinearity of the
Rashba effect competes against the Dresselhaus $k^{3}$ term, leading to
non-monotonic dependence of the peak position on the Fermi wave vector. The
values of the spin lifetime obtained from the nonlinear Rashba model also
differ from the prediction of the linear Rashba model by up to several orders
of magnitude [cf. Figs. 4(c) and 4(d)].

In summary, we have revealed that the RSS in semiconductor QW's is
intrinsically a nonlinear function of the wave vector. It may be overestimated
significantly by the linear Rashba model, especially in narrow-gap QW's. We
propose a two-coefficient nonlinear Rashba model, which is in good agreement
with the numerical results obtained from the eight-band $\mathbf{k}%
\cdot\mathbf{p}$ theory. Using this nonlinear model, we found pronounced
suppression of the DP\ SRR at large electron density, and a non-monotonic
dependence of the ratio $(-\alpha/\beta_{D})$ at the resonance peak of the
electron spin lifetime on the Fermi wave vector in [111]-oriented QW's, both
in qualitative disagreement with the predictions of the linear Rashba model.
The values of the spin lifetime obtained from the two models may differ by up
to several orders of magnitude. Further surprising results are expected when
the nonlinear Rashba model is applied to other fields.

This work was partly supported by the NSF and MOST of China.

\end{document}